\UseRawInputEncoding

\documentclass[letter,prapplied,
 amsmath,amssymb,
 reprint,
 author-year,%
superscriptaddress,
longbibliography
]{revtex4-1}

\usepackage{natbib}
\usepackage[dvipsnames]{xcolor}
\usepackage{float}
\usepackage{graphicx}
\usepackage{tabularx}
\usepackage{epstopdf}
\AppendGraphicsExtensions{.tif}
\usepackage{amsmath,amsthm,amssymb,amsfonts}
\usepackage{dcolumn}
\usepackage{bm}
\usepackage{gensymb}
\usepackage[normalem]{ulem}

\newcolumntype{C}{>{\centering\arraybackslash}X} 

\usepackage{cleveref}
\AtBeginDocument{%
    \newwrite\bibnotes
    \def\bibnotesext{Notes.bib}
    \immediate\openout\bibnotes=\jobname\bibnotesext
    \immediate\write\bibnotes{@CONTROL{REVTEX41Control}}
    \immediate\write\bibnotes{@CONTROL{%
    apsrev41Control,author="08",editor="1",pages="1",title="0",year="1"}}
     \if@filesw
     \immediate\write\@auxout{\string\citation{apsrev41Control}}%
    \fi
}%

\crefname{equation}{Eq.}{Eqs.}
\Crefname{equation}{Equation}{Equations}
\crefname{figure}{Fig.}{Figs.}
\Crefname{figure}{Figure}{Figures}
\crefname{section}{Sect.}{Sects.}
\Crefname{section}{Section}{Sections}
\crefname{table}{Table}{Tables}
\crefname{appsec}{Appendix}{Appendices}
\graphicspath{{Figures/}}
\begin{document}

\title{Phase Diffusion in Low-$E_J$ Josephson Junctions at milli-Kelvin Temperatures}

\author{Wen-Sen Lu}
\author{Konstantin Kalashnikov}
\author{Plamen Kamenov}
\author{Thomas J. DiNapoli}
\author{Michael E. Gershenson}
\affiliation{Department of Physics and Astronomy, Rutgers University, Piscataway, NJ}

\date{\today}

\begin{abstract}

Josephson junctions (JJs) with Josephson energy $E_J \lesssim 1K$  are widely employed as non-linear elements in superconducting circuits for quantum computing, operating at milli-Kelvin temperatures. Here we experimentally study incoherent phase slips (IPS) in low-$E_J$ Aluminum-based JJs at $T<0.2K$, where the IPS become the dominant source of dissipation. We observed strong suppression of the critical (switching) current and a very rapid growth of the zero-bias resistance with decreasing Josephson energy below $E_J \sim 1K$. This behavior is attributed to the IPSs whose rate exponentially increases with decreasing the ratio $E_J/T$. Our observations are in line with other data reported in literature. With further improvement of coherence of superconducting qubits, the observed dissipation from IPS might limit the performance of qubits based on low-$E_J$ junctions. Our results point the way to future improvements of such qubits.

\end{abstract}

\pacs{Valid PACS appear here}
\keywords{Suggested keywords}
\maketitle

\section{Introduction}

Josephson junctions (JJs) with the Josephson energy $0.1K<E_J<1K$ have been recently employed as non-linear elements of superconducting qubits (see, e.g., \cite{Nguyen2019-xc,Gyenis2021-ef,Zhang2020-vr,Peruzzo2020-ds}). Though $E_J$ of these junctions remains much greater than the physical temperature of qubits ($\sim 20 \div 50$ mK), a non-zero rate of thermally activated phase slips in these junctions might soon limit the coherence of superconducting qubits. Indeed, with the qubit coherence time exceeding 1 ms \cite{Somoroff2021-ep}, even rare dissipative events might become significant. Thus, the study of incoherent phase slips, induced by either equilibrium (thermal) or non-equilibrium noise, might help better understand the limitations of the low-$E_J$ JJs as elements of quantum circuits operating at mK temperatures. 

In the past, phase slips in JJs \cite{Tinkham1996-db} and associated phase diffusion \cite{,Martinis1989-rn,Kautz1990-pj,Eiles1994-nj,Watanabe2003-ko} attracted a great deal of experimental and theoretical attention. This effort was mainly aimed at better understanding of a crossover from the classical Josephson behavior (well-defined phase difference, strong quantum fluctuations of charge) to the Coulomb-blockade regime (localized charges, strong quantum fluctuations of phase) (see, e.g., \cite{Fazio2001-lo,Bard2017-ad,Ast2016-gn,Murani2020-zr} and references therein). The crossover is observed in ultra-small JJs with the Josephson energy $E_J$ of the same order of magnitude as the Coulomb energy $E_C=(2e)^2/(2C_J )$ ($C_J$ is the effective JJ capacitance) provided the junctions are included in a circuit with the impedance $Z$ greatly exceeding the quantum resistance $R_Q=h/(2e)^2 \approx 6.5k\Omega$. The rate of the coherent phase slip processes (the so-called quantum phase slips, or QPS) exponentially increases with decreasing the ratio $E_J/E_C$ \cite{Matveev2002-yb}. QPS might induce the qubit dephasing \cite{Manucharyan2012-cb} in the long-coherence superconducting qubits such as transmons \cite{Koch2007-ni} and “heavy fluxoniums” \cite{Nguyen2019-xc,Gyenis2021-oy,Earnest2018-tc}.

In this paper we are concerned with phase slips in the regime $\Delta \gg E_J \gtrsim T \gg E_C$, where the quantum fluctuations of charge are strongly enhanced. This regime, less explored using DC measurements, is relevant for operation of long-coherence superconducting qubits shunted with a large external capacitance \cite{Ast2016-gn,Murani2020-zr,Matveev2002-yb}. To explore the dynamics of low-$E_J$ junctions at mK temperatures, we designed JJs with $E_J=0.1-1K$ and $E_C<10$ mK, and studied the dissipative processes in these JJs in low-frequency transport measurements.
The paper is organized as follows. In Section II we briefly review the known facts about the phase diffusion induced by incoherent phase slips in underdamped JJs. The sample design and experimental techniques are discussed in Section III. The measurements of current-voltage characteristics (IVC) of low-$E_J$ devices are presented in Section IV. In Section V we discuss the results, compare them with the data reported by other experimental groups, and consider the implications of the dissipation induced by incoherent phase slips for the operation of qubits that employ low-$E_J$ Josephson junctions.  We provide our conclusions in Sec. VI.

\section{Phase Diffusion in Underdamped Junctions}

At $T=0$, the critical current $I_C^{AB}$ of a “classical” JJ ($E_J \gg E_C$) is provided by the Ambegaokar-Baratoff relation \cite{Tinkham1996-db}

\begin{equation}
    I_C^{AB} (T=0)=\frac{2e}{\hbar} E_J = \frac{\pi \Delta(0)}{2eR_N},
\end{equation}

where $\Delta$ is the superconducting energy gap and $R_N$ is the normal-state resistance of a JJ.  This relation has been derived by neglecting phase fluctuations. In the absence of non-equilibrium noise and charging effects, the voltage drop across a JJ is expected to be zero at $I<I_C^{AB}(T=0)$. 
The quantum phase fluctuations, which become strong at $E_J\lesssim E_C$, result in the so-called coherent quantum phase slips (CPS) in one-dimensional JJ chains (see \cite{Mooij2006-vq,Astafiev2012-ih,Svetogorov2018-ex} and references therein). The junction capacitance $C$ plays the role of the effective mass of a fictitious particle that tunnels between the minima of the “washboard” potential $U(\varphi)=-E_J \cos{\varphi} - \frac{\hbar I}{2e} \varphi $ \cite{Tinkham1996-db}. Reduction of $C$ and, thus, increase of $E_C$, facilitates tunneling and promotes CPS. The CPS shift the system energy levels and renormalize the effective Josephson coupling $E_J^*\sim E_J^2/E_C$, but do not lead to energy dissipation (in contrast to the incoherent quantum phase slips in one-dimensional superconducting wires \cite{Arutyunov2008-ar,Semenov2016-ny}). 

In the $E_J\gg E_C$ regime, on the other hand, incoherent classical phase slips (IPS) induced by either non-zero temperature or non-equilibrium noise are expected to dominate.  IPS correspond to the over-the-barrier activation in the washboard potential \cite{Tinkham1996-db}.  In thermal equilibrium the IPS rate depends exponentially on the temperature: $\nu_{IPS}\approx \omega_p e^{-\Delta U/k_BT}$ \cite{Tinkham1996-db}. Here $\omega_p=\frac{1}{\hbar} \sqrt{2E_J/E_C}$ is the plasma frequency which plays the role of the attempt rate, $\Delta U$ is the height of the potential barrier which is close to $2E_J$ at currents $I \ll I_C^{AB}$. 

The IPS process is analogous to a single flux quantum $\Phi_0$ crossing a JJ (the process is dual to the transfer of a single Cooper pair through the JJ \cite{Schon1990-ul}. Each phase slip generates a voltage drop $V(t)$ across the JJ, such that $\int V(t) dt=\Phi_0$ and, in the presence of a current $I$, releases an energy $I\Phi_0$. Thus, the zero-voltage state can be destroyed by the energy dissipation due to the time-dependent phase fluctuations.  At zero tilt of the “washboard” potential $U(\varphi)$, the phase slips with different signs of the phase change occur with the same probability and, as a result, the average voltage across the junction is zero. However, when the junction is biased with a non-zero current $I$, the tilt of the washboard potential breaks the symmetry and a non-zero average voltage proportional to the phase slip rate is generated across the junction. 

The dynamics of JJs depends on all sources of dissipation, such as IPS, thermally excited quasiparticles, etc. The low-dissipative (underdamped) regime, observed at $T \ll \Delta$ and in a high-impedance environment, is relevant to the operation of superconducting qubits. Typically, dissipation is highly frequency dependent: it is strongly suppressed at low frequencies $\omega \ll \omega_p$ and, potentially, significantly enhanced at frequencies approaching $\omega_p$. This frequency-dependent dissipation leads to the phenomenon of underdamped phase diffusion \cite{Martinis1989-rn,Kautz1990-pj,Vion1996-uj}. Characteristic signatures of this regime are the absence of the zero-voltage superconducting state and the existence of a low-voltage ($V\ll\frac{2\Delta}{e}$) IVC branch, which extends up to $I_{SW} \ll I_C^{AB}$. The IVC is hysteric at currents $I<I_{SW}$: the low-V branch observed with increasing the current from $0$ to $I_{SW}$ coexists with a high-voltage ($V\geq\frac{2\Delta}{e}$) branch observed with decreasing the current from $I>I_{SW}$ to zero (see Fig.2). At high voltages $V>\frac{2\Delta}{e}$ the main dissipation mechanism is the Cooper pair breaking and generation of non-equilibrium quasiparticles. In the low-volage state $V<\frac{2\Delta}{e}$ the energy gained by a system in the process of the over-the-barrier activation is dissipated mostly due to the Josephson radiation \cite{Fistul2016-th}.

The theory of the DC transport in underdamped Josephson junctions in the regime $E_C\ll T \leq E_J<\Delta$ in presence of a stochastic noise has been developed by Ivanchenko and Zilberman \cite{Ivanchenko1969-et} (the IZ theory, see Appendix 3). The IZ theory predicts that $I_{SW}\propto E_J^2$ at small $E_J$ \cite{Shimada2016-kk}, in contrast to the dependence $I_C^{AB}\propto E_J$ for the regime $E_C,T \ll E_J$ (Eq.1).

More recent analysis of the effect of non-zero temperature in the underdamped junctions was provided by Kivioja et. al. \cite{Kivioja2005-dq}. By considering the quality factor at the plasma frequency, $Q(\omega_p)$, and the energy dissipated between adjacent potential maxima $\Delta E_D\approx 8E_J⁄Q(\omega_p)$, Kivioja et. al. showed that the maximum possible power dissipated due to phase diffusion before switching to a state with $V\approx 2\Delta/e$ can be expressed as 

\begin{equation}
    \frac{2\pi V}{\Phi_0} \times \frac{\Delta E_D}{2\pi} = V \times I_{SW},
\end{equation}

where $I_{SW}=4I_C⁄\pi Q$ is the maximum possible current carried by underdamped junctions in the phase diffusion (UPD) regime. At $I<I_{SW}$, there is a non-zero probability for a fictitious particle to be retrapped after escape from a local minimum of the potential $U(\varphi)$. As a result, instead of a run-away state with $V=2\Delta⁄e$, the IVC demonstrates a non-zero slope at $I<I_{SW}$ due to the phase diffusion. The value of $R_0$, therefore, provides valuable information regarding the nature of damping in the junction circuits.

\section{Experimental techniques}

All the samples studied in this work have been implemented as SQUIDs, in order to be able to \textit{in-situ} tune $E_J$ by changing the magnetic flux $\Phi$ in the SQUID loop \cite{Tinkham1996-db}:

\begin{equation}
    E_J = 2E_{J0} \cos{(\pi \frac{\Phi}{\Phi_0})},
\end{equation}

Figure 1 schematically shows the design of a chain of SQUIDs formed by small ($0.2 \times 0.2\mu m^2$) JJs. The area of the SQUID loop, $A_{SQUID}$, varied between $6\mu m^2$ and $50\mu m^2$. The chains of SQUIDs had additional contact pads (shown in yellow in Fig. 1) to provide access to individual SQUIDs or pairs of SQUIDs within a chain.

\begin{figure}
\centering
\includegraphics[scale=0.55]{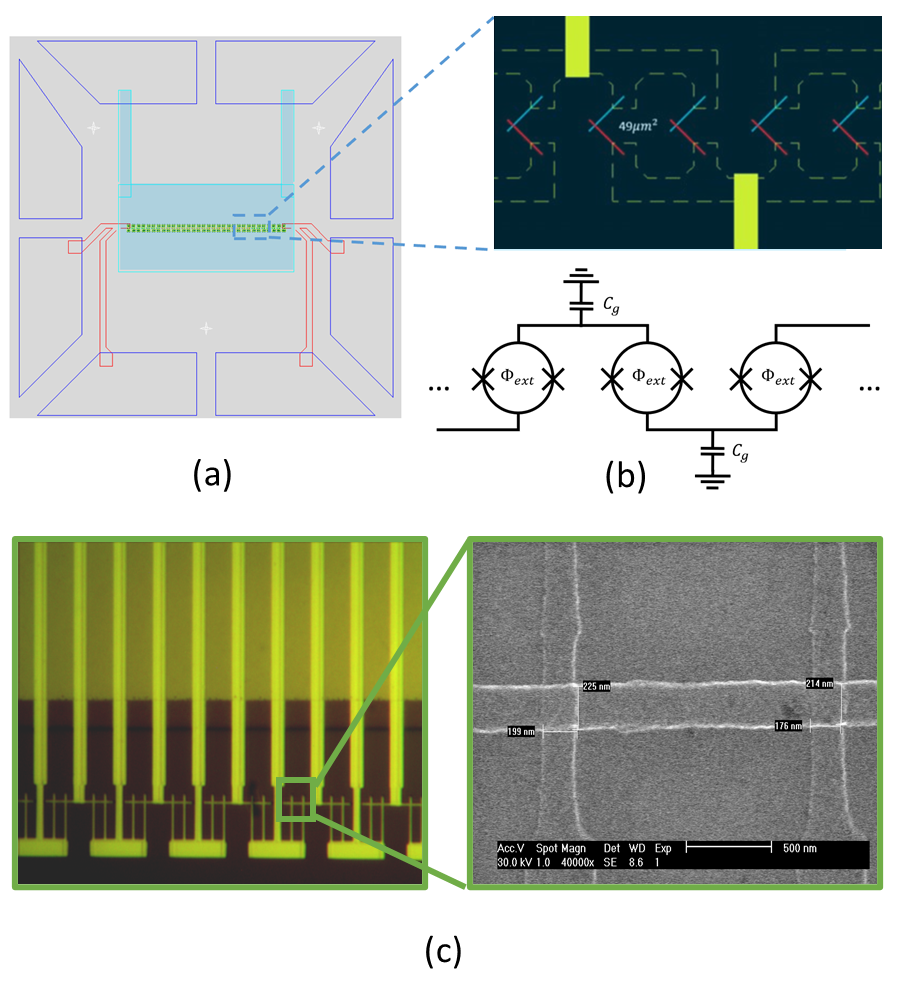}
\caption{(a) Schematics of a chain of SQUIDs made of Josephson junctions with a relatively large area (i.e. large $C_J$) and a low transparency of the tunneling barrier (i.e. small $E_J$). The common ground electrode made of a sputtered Pt film is shown in pale blue. A few-$nm$-thick $AlO_X$ oxide covers this electrode and serves as a pinhole-free dielectric that isolates the ground from the SQUIDs. The typical value of the capacitance that shunts a single SQUID, $C_g$, is $0.5nF$ for $50\mu m^2$ pad area. This $C_g$ corresponds to a charging energy per SQUID $E_C=\frac{(2e)^2}{2C_g}=8$ mK. (b) The circuit diagram of a chain of SQUIDs. (c) An alternative design of a chain of SQUIDs shunted by external capacitors to the ground. The vertical $5\mu m$-wide pads are the ground electrodes for the capacitors, a few-$nm$-thick $AlO_X$ serves as a dielectric between the electrodes.}
\label{fig:device_fab}

\end{figure}

In order to reduce the rate of quantum phase slips $\nu_{QPS} \propto \exp{-\sqrt{\frac{2E_J}{E_C}}}$ \cite{Tinkham1996-db}, we used either the low-transparency JJs junctions with a relatively large in-plane area $A_{JJ}$ ($>1\mu m^2$), or smaller junctions shunted with external capacitors (the design details are provided in Appendix 1). In both cases the charging energy $E_C$ was reduced below $\approx 10$ mK, and this allowed us to maintain a large ratio $E_J/E_C$  for all studied JJs.

The amplitude of variations of $E_J$ with the external magnetic field depends on scattering of parameters of individual JJs that form a nominally symmetric SQUID. This scattering did not exceed $10\%$ for the JJs with the normal-state resistance $R_N \approx 1k\Omega$ and $A_{JJ}=0.02\mu m^2$. However, fabrication of the low-transparency JJs with $R_N \approx 100 k\Omega$ and $A_{JJ}=4\mu m^2$ (the nominal critical current density $I_C^{AB}/A_{JJ}\approx 5\times10^{-4} A/cm^2$), which required very long oxidation times and high partial pressure of $O_2$, resulted in a larger ($\approx30\%$) scattering of the $R_N$ values (Appendix 1). This scattering was one of the reasons for different dependencies $I_{SW}(B)$ observed for the nominally identical SQUID chains (see below). The parameters of representative samples are listed in Table 1 (the total number of tested samples exceeded $50$ \cite{Lu2021-ky}).

\begin{table}

\centering
\caption{Parameters of single Josephson junctions in SQUID chains. $R_N$ and $A_{JJ}$ are the normal-state resistance and the junction area, respectively. The Josephson energy $E_J=\pi\hbar\Delta/((2e)^2 R_N)$ has been calculated using $R_N$ and $T_C=1.3$ K . The charging energy $E_C$, where $C$ is the shunting capacitance, did not exceed $10$ mK for all samples. The critical current $I_C^{AB}$ was calculated using Eq. (1).}

\rule{8.5cm}{0.4pt}

\begin{tabularx}{\textwidth}{c|ccccc}
sample & $R_N$ ($k\Omega$) & $E_J$ (K) & $A_{JJ}$ ($\mu m^2$) & $I_C^{AB}$ & $I_{SW}$ (nA) \\
1      & 2.4               & 2.9       & 1.9                  & 130        & 48          \\
2      & 2.9               & 2.4       & 3.74                 & 107        & 68          \\
3      & 9.4               & 0.76      & 0.04                 & 33         & 9           \\
4      & 15.8              & 0.45      & 0.04                 & 20         & 0.3         \\
5      & 16.6              & 0.43      & 0.04                 & 19         & 0.1         \\
6      & 175               & 0.04      & 0.04                 & 1.8        & 0.003      
\end{tabularx}
\rule{8.5cm}{0.4pt}
\label{table:sample_summary}
\end{table}

\section{Current-voltage characteristics of low-$E_J$ junctions}

Below we focus on the results obtained at $T<200$ mK – in this temperature range one can neglect transport of the thermally-excited quasiparticles in $Al$-based superconducting circuits. Typical IVC measured at $T_{base}=25$ mK for the samples with $E_J\approx 1$ K and $E_J \ll 1$ K are shown in Figure 2. Below we address several characteristic features of the IVC.

\begin{figure*}[t!]
    \centering
    \includegraphics[scale=0.7]{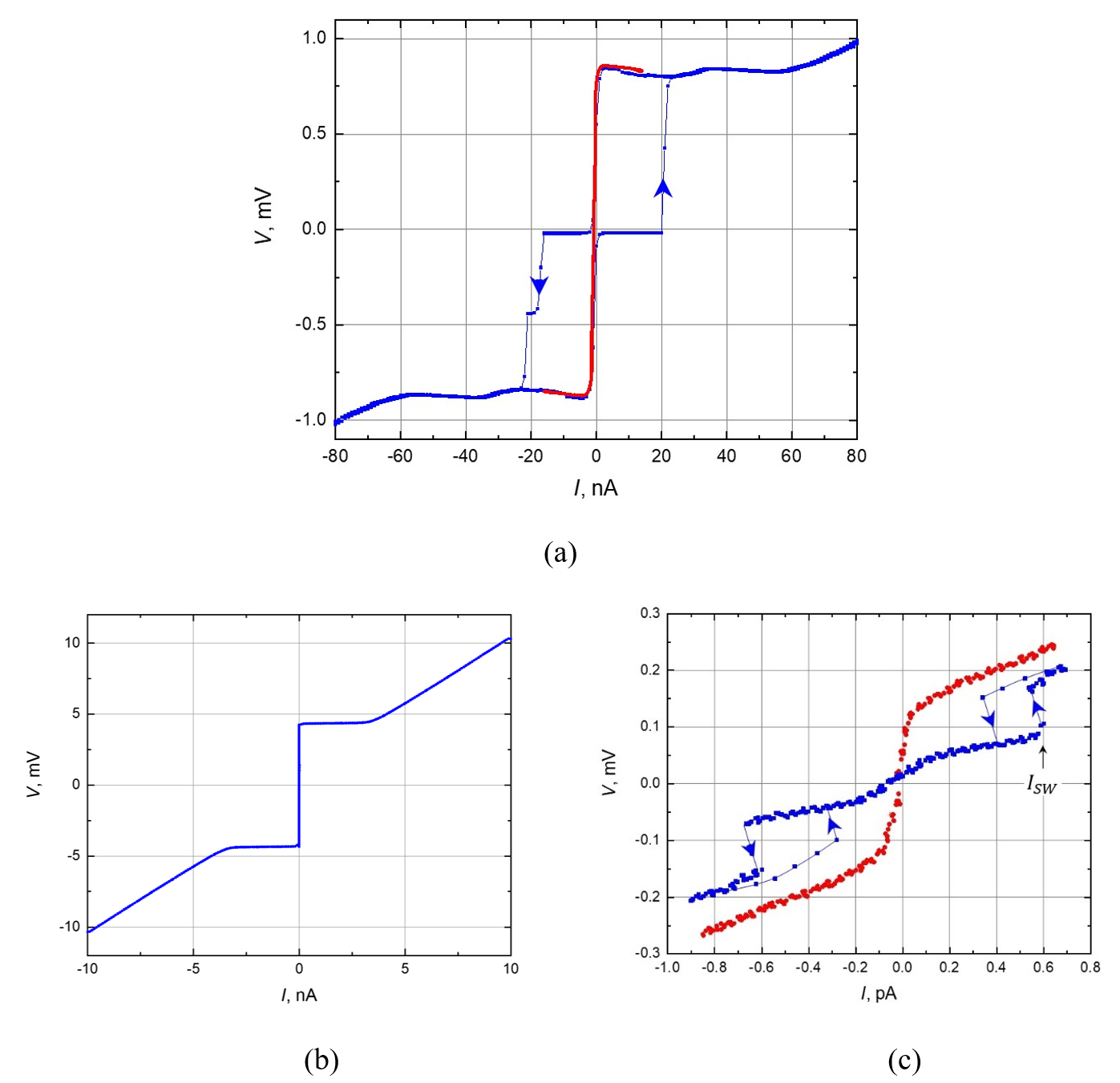}
    \caption{(a) Current-voltage characteristics of two connected-in-series SQUIDs at $\Phi=0$ (blue curve) and $\Phi=0.5\Phi_0$ (red curve) at $T\approx 30$ mK. Each SQUID is formed by two nominally identical JJs with $E_J=0.76K$ (sample 3 in Table 1), thus the SQUID Josephson energy is $1.52K$. Even for this circuit with relatively high $E_J$, the measured switching current per junction, $I_{SW}=9$ nA, is significantly lower than $I_C^{AB}=33$ nA. (b) The IVC of a chain of $20$ SQUIDs with $E_J=80$ mK (for single-JJ parameters, see sample 6 in Table 1). (c) The enlargement of the region of small currents/voltages in panel (b). Note that the resistance is non-zero for all biasing currents. The switching current (its value for a given sample, $0.6$ pA, is indicated by an arrow) corresponds to a rapid increase of the voltage across the chain. This switching current is almost four orders of magnitude smaller than the $I_C^{AB}$ value for this sample (see Table 1). The zero-bias resistance ($R_0\approx 500M\Omega$ per junction) was determined as the slope of the IVC at $I\ll I_{SW}$. }
    \label{fig:overviewIVc}
\end{figure*}

\subsection{The switching current $I_{SW}$ and the zero-bias resistance $R_0$}

Figure 2 shows how we determined the switching current $I_{SW}$ and the zero-bias resistance $R_0$ measured at small DC voltages $V\ll 2\Delta/e$ and currents $I\ll I_{SW}$. Note that the zero-bias resistance per junction is twice as large as the zero-bias resistance of a SQUID. For the JJs with $E_J=0.76$ K (Fig. 2a) a non-zero $R_0$ could not be detected within the accuracy of our measurements ($\approx 10^2 \sim 10^3 \Omega$, depending on the magnitude of $I_{SW}$).  This is the behavior expected in the “classical” regime $E_J\gg T,E_C$. At currents $I>I_{SW}$, the voltage across the chain approaches the value $N\times 2\Delta/e$, where $N$ is the number of SQUIDs in the chain and $2\Delta \approx e \times 0.4mV$ is the sum of superconducting energy gaps in the electrodes that form a junction. For the chains with $E_J\ll 1$ K, the switching current is several orders of magnitude smaller than the Ambegaokar-Baratoff critical current (Fig. 2c).
With the magnetic field $B$ approaching the value $\Phi_0/(2A_{SQUID})$ , the switching current vanishes and $R_0$ increases by orders of magnitude (red curves in Figs. 2 a,c ). The IVC at $\Phi=\Phi_0/2$ resemble that observed in the Coulomb-blockade regime. Note that for most of the studied samples in this regime $E_C$ is close to the base temperature, so the Coulomb blockade is partially suppressed by thermal effects. The resistance $R_0(\Phi=\Phi_0/2)$ for the samples with $E_J\ll 1$ K is limited by the input resistance of the preamplifier (a few $G\Omega$).

The evolution of the IVC measured at different temperatures for $\Phi=\Phi_0/2$  is shown in Fig. 3a. The $R_0(T)$ drop observed with an increase of temperature at $T>0.2$ K (Fig. 3b) is due to an increasing concentration of thermally excited quasiparticles in $Al$ electrodes: the JJ becomes “shunted” by the quasiparticle current. The dependence $R_0(T)$ at $T>0.25$ K can be approximated by the Arrhenius dependence $R_0(\Phi=\Phi_0/2,T)\propto exp(\delta/(k_B T))$ with $\delta\approx 2.1$ K. The activation energy $\delta$ is close to the superconducting energy gap $\Delta \approx 2.3$ K in $Al$ electrodes with $T_C\approx1.3$ K. A weak decrease of $R_0$ with decreasing $T$ has been observed at $T<0.2$ K for most of the studied samples; this decrease was less pronounced than the one observed in Ref. \cite{Martinis1989-rn}. 

\subsection{The IVC hysteresis}

For all studied samples we observed strong hysteresis of the IVC at $\Phi=n\Phi_0$ where $n$ is integer. The hysteresis is a signature of the underdamped junctions with the McCumber parameter $\beta \gg 1$ \cite{McCumber1968-hf}. Observation of the hysteresis is also an indication that the noise currents $I_N$ in the measuring setup are significantly smaller than the switching current even for the samples with $I_{SW}$ in the sub-pico-$A$ range (in the opposite limit, $I_N>I_{SW}$, the hysteresis vanishes, see Appendix 2 and Ref.\nocite{Coon1965-nv,Wilen2021-nx} \cite{Schmidlin_undated-ri}). 

\subsection{The $I_{SW}(B)$ dependences}

\begin{figure*}[t!]
    \centering
    \includegraphics[scale=0.7]{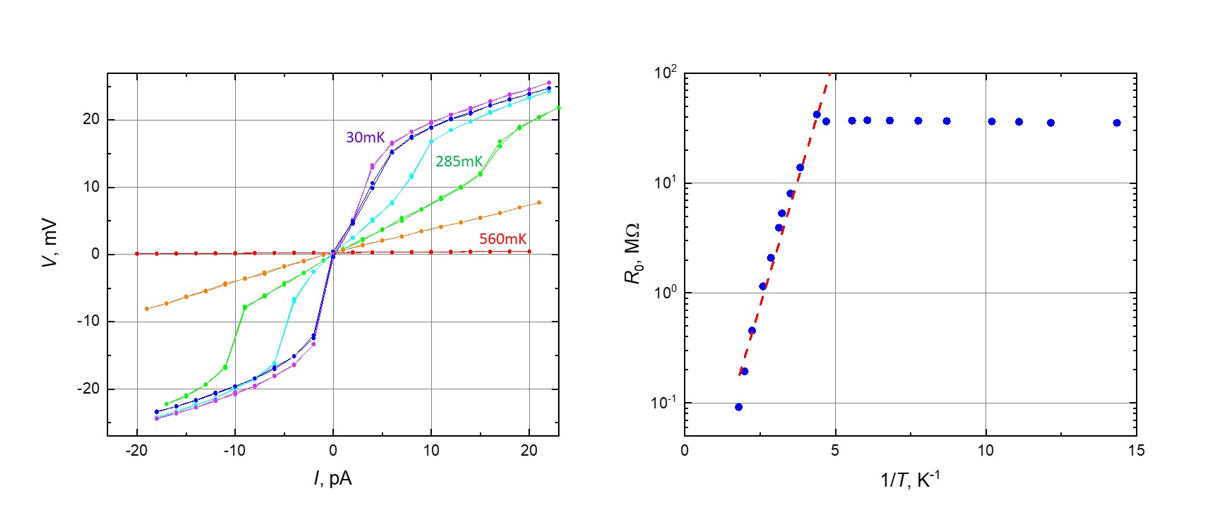}
    \caption{(a) Current-voltage characteristics of two connected in series SQUIDs measured at $\Phi=0.5\Phi_0$ and different temperatures (from $30$ mK to $560$ mK, as shown in the panel). The SQUIDs are formed by JJs with $E_J=0.76$K (sample 3 in Table 1). (b) The temperature dependence of the zero-bias resistance for this sample. The red dashed line corresponds to the dependence $R_0$  ($\Phi = 0.5 \Phi_0,T)=4k\Omega \times \exp({\delta/T}$) with $\delta=2.1$ K.}
    \label{fig:TBdep}
\end{figure*}

\begin{figure*}[t]
    \centering
    \includegraphics[scale=0.8]{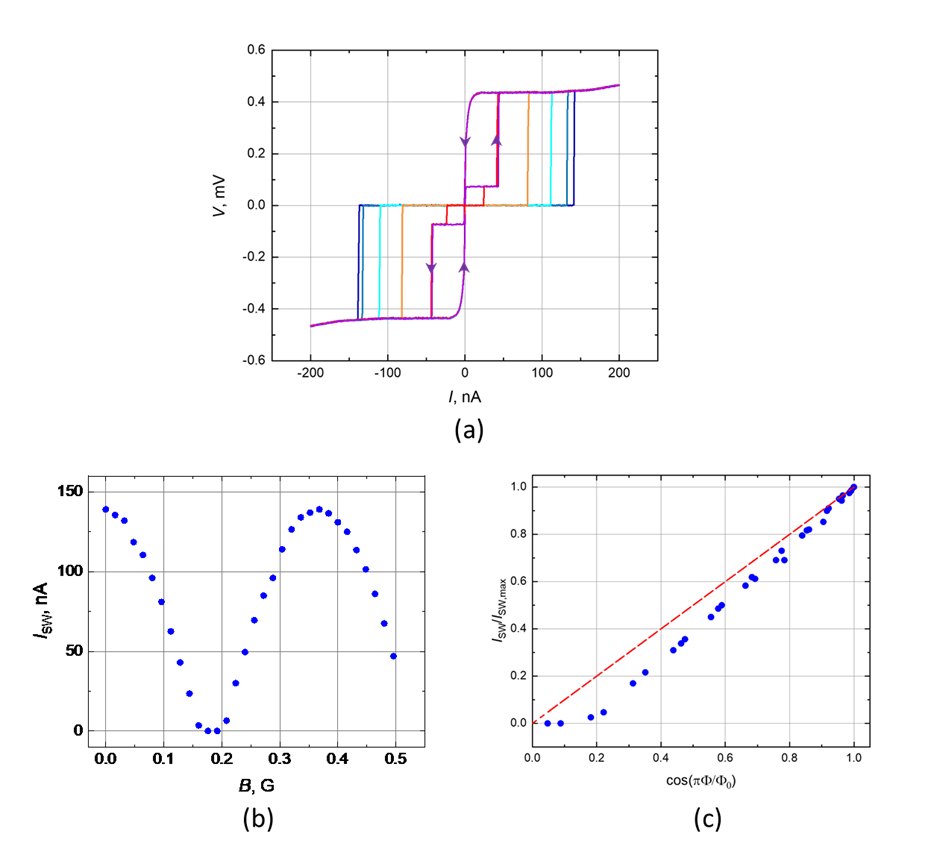}
    \caption{(a) Current-voltage characteristics of a single SQUID formed by JJs with $E_J=2.4$ K (sample 2 in Table 1) measured at different values of $\Phi/\Phi_0 = 0, 0.08, 0.17, 0.25, 0.37, 0.5$. A sub-gap voltage plateau at $V\approx75\mu V$ appears at $\Phi>0.35\Phi_0$. For different samples the sub-gap voltage plateau was observed at $V=40\sim 200\mu V$. (b) The dependence of $I_{SW}$ on the magnetic field $B$. (c) The measured $I_{SW}(\Phi)/I_{SW}(\Phi=0)$ as a function of $cos(\pi\Phi/\Phi_0)$. The dash line corresponds to the dependence $I_{SW}\propto cos(\pi\Phi/\Phi_0)$.}
    \label{fig:fluxdep}
\end{figure*}

Even in the “classical” regime $E_J \gg E_C,T$ we obtained several unexpected results. Firstly, the dependence of $I_{SW}(\Phi)$ for some samples significantly deviated from the dependence

\begin{equation}
    I_{SW}(\Phi)=I_{SW}(\Phi=0) \times \cos(\pi\frac{\Phi}{\Phi_0}),
\end{equation}

(see Fig. 4c). These deviations can be at least partially explained by a relatively large scattering of parameters of individual JJs and non-uniformity of the local magnetic field in the SQUID loops due to the magnetic field focusing. Observation of a steeper drop of $I_{SW}$ with $\Phi\rightarrow0.5\Phi_0$ than that predicted by Eq. 3 can be attributed to violation of the condition $E_J(\Phi) \gg E_C$ and crossover to the Coulomb blockade regime.

Secondly, we have observed sub-gap ($V_{subgap}<2\Delta/e$) voltage steps on the IVC (Fig. 4a), which significantly reduced the accuracy of extraction of $I_{SW}$ and $R_0$ at the values of $\Phi$ close to $\Phi_0/2$. A possible reason for appearance of sub-gap steps might be the Fiske resonances due to the microwave resonant modes of the circuit \cite{Coon1965-nv}. Identifying the circuit elements that would be responsible for the corresponding resonance frequencies at $f_{res}=(75\mu eV)⁄h \approx 18GHz$ (this frequency corresponds to a wavelength $\sim 2.5mm$ for the electromagnetic wave propagating along the interface between a silicon substrate and vacuum) requires further investigation.

\section{Discussion}

There are several potential sources of dissipation in Josephson circuits at $T\ll\Delta$, such as non-equilibrium quasiparticles or two-level systems in the circuit environment (see, e.g., \cite{Wilen2021-nx} and references therein). However, we are unaware of a mechanism other than the IPS that would explain the observed strong dependence of dissipation on the ratio $E_J/T$. Below we focus on the IPS as the dominant dissipation mechanism in our experiments.

\subsection{The switching currents $I_{SW}$}

In the regime $E_C\ll T \leq E_J<\Delta$ IPS are induced by the voltage noise - either equilibrium noise generated by thermal excitation or a non-equilibrium noise. The theory of the DC transport in underdamped Josephson junctions with $E_C\ll E_J$ has been developed by Ivanchenko and Zilberman \cite{Ivanchenko1969-et}. According to the IZ theory, a voltage-biased Josephson junction is subject to thermal noise of the biasing resistor which causes a phase diffusion. The equation for the phase $\varphi$ across a classical Josephson junction ($E_C\ll E_J$) can be written as:

\begin{equation}
    \frac{2e}{\hbar}(I_b+I_n) = \frac{1}{R}\frac{\partial\varphi}{\partial t}+\frac{2e}{\hbar}I_C sin\varphi,
\end{equation}

where $I_b$ is the bias current, $\langle I_n(0)I_n(\tau)\rangle\geq\frac{2k_B T}{R}\delta(\tau)$ is the delta-correlated Johnson-Nyquist noise across the resistance $R$ connected in parallel with the junction. By solving the corresponding Fokker-Planck equation, the superconducting part of the current as a function of bias voltage $V_B$ can be found (see Appendix 3). 
Two features of the IZ theory should be noted. First, the theory predicts quadratic drop of the maximum superconducting current that the junction can sustain (i.e. the switching current $I_{SW}$) with decreasing $E_J$ at small $E_J$ \cite{Shimada2016-kk}. This is in contrast to the Ambegaokar-Baratoff (AB) critical current $I_C^{AB}$ which decreases proportional to $E_J$. Second, the maximum value of the switching current is realized at a non-zero voltage $V_n$, which depends only on the voltage noise amplitude, so the zero-bias resistance in the IPS regime is expected to scale as $E_J^{-2}$.
Indeed, the observed dependences $ln(I_S)$ vs. $ln(E_J)$ are steeper than the linear dependence $I_{SW}(E_J )$ predicted by the Ambegaokar-Baratoff relationship (Eq. 1). For comparison, we plotted on the same plot the values of $I_{SW}$ reported by several experimental groups. Note that data from the literature correspond to samples with different $E_C$ (the ratio $E_J/E_C$ for a given $E_J$ varies over a wide range, see Table 2). This might be one of the reasons for a strong scattering of $I_{SW}$ at a given $E_J$.

\begin{figure*}[t!]
    \centering
    \includegraphics[scale=0.9]{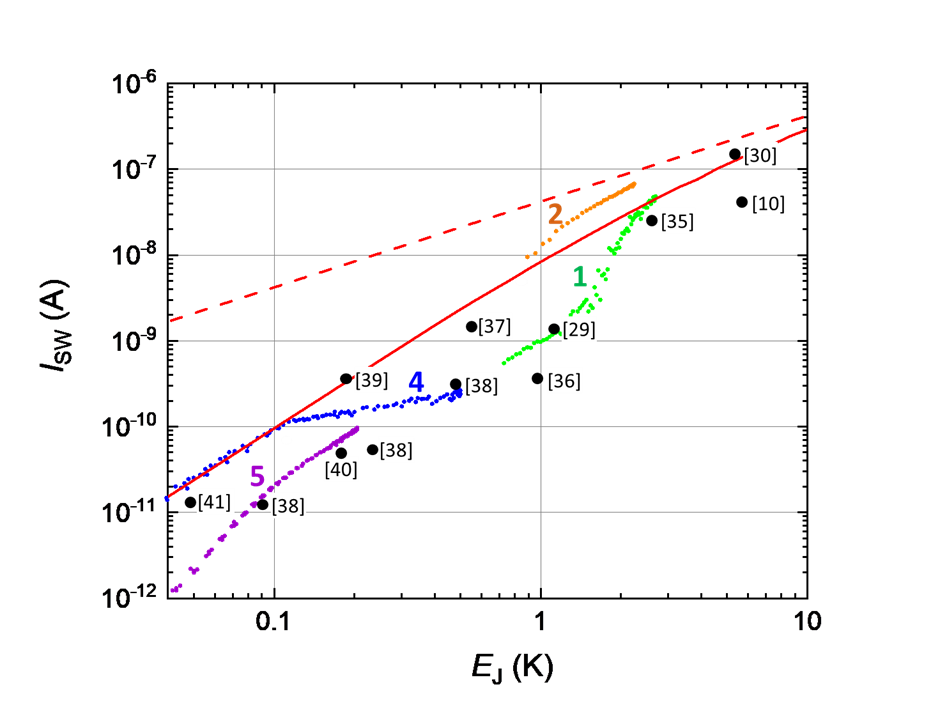}
    \caption{The switching current $I_{SW}$ as a function of $E_J$ measured in our experiments (the color-coded symbols, the sample numbers correspond to that in Table 1) and by other experimental groups (black dots, the references are given in square brackets). All the data have been obtained at $T\approx 20-50$ mK for $Al/AlO_X/Al$ junctions. For the values of $I_{SW}$ measured at $B \neq 0$ the Josephson energy $E_J (B)$ was calculated using Eq. (2). The dashed red line corresponds to the Ambegaokar-Baratoff dependence $I_C^{AB}(E_J)$ (Eq. 1), the solid red curve - to the switching current predicted by the IZ theory in presence of additional $V_{noise} = 20\mu V$  generated by the biasing scheme (see Appendix 4).}
    \label{fig:IswEJ}
\end{figure*}

In Fig. 5 we plotted $I_{SW}(E_J)$ predicted by the IZ theory in presence of additional Gaussian noise with amplitude of $V_{noise}=20\mu V$. This noise corresponds to the Johnson-Nyquist noise $\delta V_t=\sqrt{4k_B TR\Delta f}$ generated at $T=50$ mK by two $100$ $k\Omega$ resistors connected in series with the device (Fig. A2). These chip resistors, designed for microwave applications, had a very small imaginary part of their impedance. The bandwidth was estimated as $\Delta f\approx \omega_p⁄2\pi$, where $\omega_p⁄2\pi\approx 1GHz$ is the plasma frequency of the shunted JJs. Most of the $I_{SW}$ data points in Fig. 5 are still 1-2 orders of magnitude smaller than $I_{SW}$  predicted by the IZ theory. A possible explanation for this discrepancy might be more complex phase dynamics in the devices with a very high IPS rate, outside of the limits of applicability of the IZ theory. Another possibility is the exponentially strong sensitivity of the IPS rate to the noise level in different setups and the physical temperature of samples, the parameters that are not easy to control in most experiments.
Figure 5 shows the data for four samples whose $E_J$ was varied by the external magnetic flux threading the SQUIDs loop. The effective $E_J$ for these devices was calculated using Eq. 2. By tuning $E_J$ over an order of magnitude, we observed rather complicated dependences $I_{SW}(E_J)$ that varied between $\sqrt{E_J}$ and $E_J^2$.

\begin{figure*}[t!]
    \centering
    \includegraphics[scale=1]{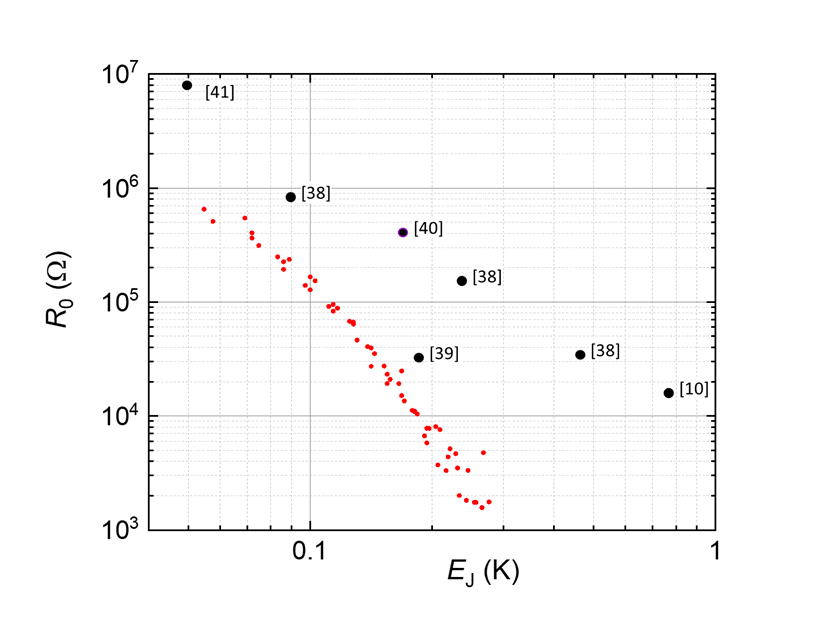}
    \caption{The zero-bias resistance $R_0$ as a function of $E_J$ measured for a chain of SQUIDs made of JJs with $E_J=0.43$ K (sample 5 in Table 1) (red dots). For comparison, we also plot the values of $R_0$ measured by other experimental groups for $Al/AlO_X/Al$ junctions (black dots, the references are given in square brackets), the parameters of these samples are listed in Table 2. All the data have been obtained at the base $T<50$ mK, though the physical temperature of the Josephson circuits has not been directly measured. The Josephson energy $E_J(B)$ for sample was calculated using Eq. (2).}
    \label{fig:IVcB}
\end{figure*}

\subsection{The zero-bias resistance $R_0$}

Figure 6 shows $R_0$  as a function of $E_J$ measured in our experiments and by other experimental groups for $Al/AlO_X/Al$ junctions. To simplify the Figure, we plotted our data only for the sample with the lowest values of $R_0$ (sample 5); $R_0$ for other samples are approximately in line with the data from literature shown in Fig. 6. The zero-bias resistance, being unmeasurably low at $E_J>1$ K, rapidly increases at $E_J<1$ K, and becomes much greater than the normal-state resistance $R_N$ at $E_J\leq0.1$ K. Instead of a well-defined “superconductor-to-insulator” transition at a certain value of $E_J/E_C$, a broad crossover between these two limiting regimes is observed. Note that different JJ samples (single junctions and arrays) demonstrate similar values of $R_0$, though their charging energies could vary over a wide range.

Figures 5 and 6 show that our findings are in good agreement with the literature data on the highest values of $I_{SW}$ and lowest values of $R_0$ measured for low-$E_J$ junctions. Despite large scattering of the data in Figs. 5 and 6, a very rapid drop of $I_{SW}$ and increase of $R_0$ has been observed in most of the experiments as soon as $E_J$ becomes significantly less than $1$ K. Figure 5 shows that for typical experimental conditions, the crossover between the “classical” behavior $I_C \propto E_J$ to the behavior controlled by the phase diffusion occurs at $E_J\approx 1$ K. Note that the literature data in Figs. 5 and 6 correspond to samples with different values of the ratio $E_J/E_C$. However, large scattering range of $I_{SW}$ and $R_0$ hides possible effect of charging. For the same reason, it is unclear if the impedance of the environment plays any significant role in these experiments: similar values of $I_{SW}$ could be observed for single JJ in a highly-resistive environment ($>100k\Omega$ as in \cite{Weisl2015-sk} and our setup), single JJ in a low-impedance environment \nocite{Jack2015-qa}
\cite{Senkpiel2020-zf}, and chains of SQUIDs frustrated by the magnetic field \cite{Lu2021-ky,Wilen2021-nx}.

Our observations are in line with an expected strong dependence of the IPS rate on the sample parameters in the regime $E_C\ll T \leq E_J\ll\Delta$. At $E_J\gg T$, one can estimate the rate of the thermally-generated IPS as $\Gamma=\omega_p exp(-2E_J/k_B T)$, where $\omega_p$ is the plasma frequency (or an attempt rate) and $exp(-2E_J/k_B T)$ is the probability of the over-the-barrier excitation. For example, at $E_J=0.25$ K and $\omega_p/2\pi=1.32$ GHz,  the rate decreases from $3\times10^5 s^{-1}$ to $0.1 s^{-1}$ if the physical temperature decreases from $50$ mK to $20$ mK. This might also explain why the experimental results are so sensitive to the noise level in the experimental setup.

\begin{table*}[t!]

\centering
\caption{The literature data on $I_{SW}$ and $R_0$, ranked by $E_J$}
\rule{14cm}{0.4pt}
\begin{tabular}{l|ccccc}
Reference                                           & $E_J$ (K) & $E_J/E_C$ & $I_C^{AB}$ (nA) & $I_{SW}$ (nA) & $R_0$ ($k\Omega$) \\ \hline
Watanabe 2003 {[}10{]} sample C                     & 5.7       & 8         & 240             & 40            & 0.6               \\ \hline
Kivioja 2005 {[}30{]}                               & 5.2       & 500       & 220             & 145           &                   \\ \hline
Schmidlin 2013 {[}35{]}, fig. 5.2                   & 2.5       & 50        & 106             & 25            & 0.13              \\ \hline
Shimada 2016 {[}29{]}, SQUID at $\Phi/\Phi_0=0.375$ & 1.1       & 14        & 47              & 1.2           & 0.11              \\ \hline
Weissl 2015 {[}36{]} SQUID at $\Phi/\Phi_0=0.26$    & 0.95      & 10        & 38              & 0.35          & 0.14              \\ \hline
Watanabe 2003 {[}10{]} sample G                     & 0.76      & 1         & 32              &               & 14                \\ \hline
Jäck 2015 {[}37{]} fig.4.6                          & 0.54      &           &                 & 1.5           & 13                \\ \hline
Senkpiel 2020 {[}38{]}                              & 0.47      &           &                 & 0.3           & 33                \\ \hline
Senkpiel 2020 {[}38{]}                              & 0.23      &           &                 & 0.07          & 143               \\ \hline
Yeh 2012 {[}39{]}                                   & 0.18      & 1.3       & 6.5             & 0.35          & 31                \\ \hline
Jäck 2017 {[}40{]}                                  & 0.17      &           & 7..5            & 0.05          & 400               \\ \hline
Murani 2020 {[}14{]}                                & 0.12      &           & 5               & 0.07          &                   \\ \hline
Senkpiel 2020 {[}38{]}                              & 0.09      &           &                 & 0.012         & 830               \\ \hline
Kuzmin 1991 {[}41{]}                                & 0.05      & $\ll 1$    &                 & 0.014         & 8000              \\
\end{tabular}
\rule{14cm}{0.4pt}
\label{table:ref_summary}
\end{table*}

\section{Conclusion and outlook}

Phase slips in JJs have been actively studied over the last three decades in different types of Josephson circuits (single JJs, JJ arrays, etc.) over wide ranges of $E_J$ and $E_C$. In our work we focused on the incoherent phase slips, which, in contrast to the coherent quantum phase slips, result in dissipation. At sufficiently low temperatures $T\ll\Delta$, where the concentration of quasiparticles becomes negligibly low, the IPS are expected to be a significant source of dissipation.

We observed that in all studied devices with $E_J<1$ K the switching current $I_{SW}$ is significantly suppressed with respect to $I_C^{AB}$. At the same time, we observed a very rapid growth of $R_0$ with decreasing Josephson coupling below $E_J\approx 1$ K. Large scattering of the data might reflect a steep dependence of the rate of incoherent phase slips on the physical temperature and non-equilibrium noise in different experimental setups. Our observations are consistent with similar data that has been previously reported in the literature. 

\nocite{Yeh2012-eg}
\nocite{Jack2017-wx}
\nocite{Kuzmin1991-qi}
The observed enhanced dissipation in Josephson circuits with $E_J<1$ K might impose limitations on the further progress of superconducting qubits based on low-$E_J$ junctions. This important issue requires further theoretical and experimental studies. Especially important direction would be measurements of the coherence time in the qubits with systematically varied Josephson energy over the range $E_J=0.1-1$ K. One of the signatures of IPS-induced decoherence might be an observation of a steep temperature dependence of the coherence time at $T<100$ mK \cite{Manucharyan2020-ax}.
\nocite{Bell2012-ox}
\nocite{Martinis1987-mh}
\begin{acknowledgments}
We would like to thank Srivatsan Chakram for insightful discussions. The work at Rutgers University was supported by the NSF awards DMR-1708954, DMR-1838979, and the ARO award W911NF-17-C-0024.
\end{acknowledgments}

\section{References}
%

\newpage

\end{document}


\title{Supplemental Material: Phase Diffusion in Low-$E_J$ Josephson Junctions at milli-Kelvin Temperatures}

\maketitle

\setcounter{equation}{0}
\setcounter{figure}{0}
\setcounter{table}{0}
\setcounter{section}{0}
\setcounter{page}{1}
\makeatletter
\renewcommand{\theequation}{A\arabic{equation}}
\renewcommand{\thefigure}{A\arabic{figure}}
\renewcommand{\thesection}{A\arabic{section}}
\renewcommand{\bibnumfmt}[1]{[A#1]}
\renewcommand{\citenumfont}[1]{A#1}

\section{Device design and fabrication}

The Josephson junctions in this work were fabricated by the Manhattan pattern technique with multi-angle deposition of Al electrodes trough bilayer e-beam resist mask \cite{Bell2012-ox}. The oxidation process performed between deposition of the bottom and top aluminum electrodes has been optimized for fabrication of junctions with required values of $E_J$  and minimal scattering of junction parameters.  Typically, we used the dry Oxygen partial pressure $1-100$ torr and oxidized the structures for $5-15$ minutes. The standard deviation of the normal state resistance $R_N$ across the $7 \mathrm{mm} \times 7 \mathrm{mm}$ chip did not exceed $10\%$ for sub-μm-wide junctions with $R_N \sim 1\mathrm{k\Omega}$ and $30\%$ for the junctions with $R_N \sim 100 \mathrm{k\Omega}$. The junction area variations did not exceed $10\%$ across a $200\mathrm{\mu m}$-long chain.

All the samples studied in this work have been implemented as SQUIDs, in order to be able to \textit {in-situ} tune $E_J$ by applying the external magnetic field. Figure A1 schematically shows the design of a chain of SQUIDs formed by small junctions ($0.2\times0.2 \mu m^2$). The area of the SQUID loop varied between $6\mu m^2$ and $49\mu m^2$. Our experiments were focused on the JJs with $1$K $> E_J \gg E_C$: this regime is relevant to the quantum circuits in which JJs are shunted with large external capacitors (such as the transmon qubit). Large $E_J/E_C$ ratio also significantly reduces the rate of quantum phase slips $\Gamma_{QPS} \propto exp(-2\sqrt{E_J/E_C})$ \cite{Ast2016-gn}. The specific capacitance of the junction tunneling $AlO_X$ barrier is about $50 fF/\mu m^2$, and in order to reduce $E_C$ down to $\sim 10$ mK the junctions should either have relatively large in-plane dimensions ($A_{JJ} > 4 \mu m^2$) or be shunted with external capacitors ($C_g>200$fF). We have used both methods in different structures. In the approach where we introduced relatively large JJs in order to keep $E_J$ below $1$K, the oxidation recipes were fine-tuned for the growth of low-transparency tunneling $AlO_X$ barrier. In the “external capacitor” approach, several designs of the shunting capacitors have been implemented. Figure A1 shows that each SQUID unit cell is flanked by two large metal pads, which are used as shunting capacitors $C_g$ to the common ground when the entire chain was covered by an additional top electrode (sputtered $Pt$ film).  A few nm native $AlO_X$ oxide grown at the atmospheric pressure serves as a pinhole-free dielectric for this parallel-plate $C_g$ with a typical capacitance around $500$ fF for $50\mu m^2$ pad area. Such $C_g$ corresponds to a charging energy per each cell as low as $E_C=(2e)^2/2C=8$mK. 

\begin{figure}
    \centering
    \includegraphics{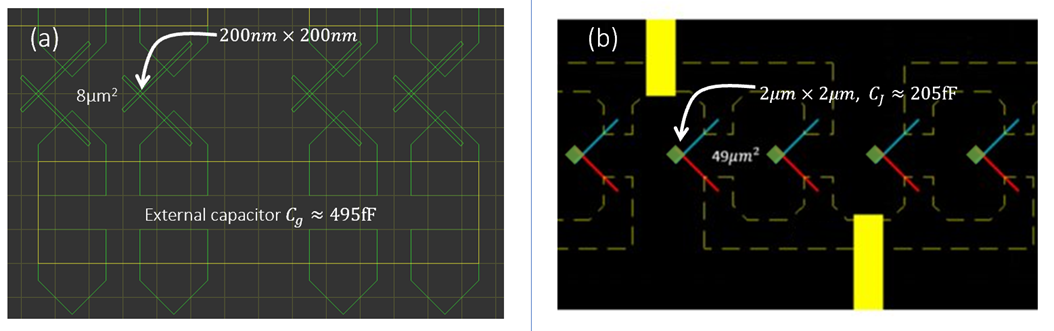}
    \caption{Various designs of SQUIDs. (a) Each SQUID unit cell was shunted by a large $\mathbf{C_g \approx 0.5}$\textbf{pF} the ground. The common ground electrode is shown by a yellow rectangle. (b) SQUIDs formed by large JJs with junction area $\mathbf{A_{JJ}\approx 2.2}$ $\mathrm{\mathbf{\mu m^2}}$. Yellow rectangles show electrodes used to measure the IVC of individual SQUIDs.}
    \label{fig:varDesign}
\end{figure}

\section{Measurement setup}

To measure the IVC of low-$E_J$ junctions with small switching currents (typically, within the pA-nA range), careful filtering of noise in the measurement circuit is required (see, e.g., \cite{Schon1990-ul}). Our measurement setup included the cascaded low pass filters shown in Fig. A2. The wiring for DC setup inside the cryostat consists of 12 twisted pairs made of resistive alloys $CuNi$:$NbTi$ (5:1) with multiple thermal anchoring points. Near the cold finger which supported the sample holder, about 1-meter-long twisted pairs are used as central conductors of the copper-powder-epoxy lowpass filter for the cut-off frequency $\sim 100$ MHz (see, e.g., \cite{Martinis1987-mh}); this filter also provides the thermal anchoring of all wires before connecting to the sample. On the sample holder, $100 k\Omega$ surface mount metal-film resistors with low parasitic capacitance have been installed in each lead. The voltage across the sample was amplified by a preamplifier (DL Instrument 1201) with a few-G$\Omega$ input impedance. 

The circuit outside of the dilution refrigerator (Fig. A2) included a commercial LC low pass filter (BLP 1.9+, DC$-1.9$ MHz) and a homemade RC filter (DC$-8$Hz) box with variable biasing resistors up to 1G$\Omega$. The voltage drop across the sample was amplified with a voltage preamp DL1201 and measured by HP 34401A digital multimeter. 

\begin{figure}
    \centering
    \includegraphics{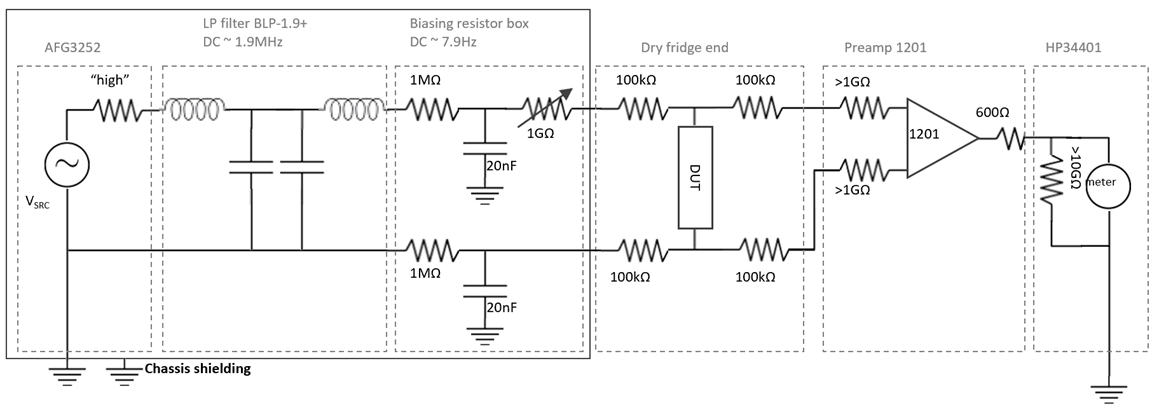}
    \caption{The wiring schematics for DC current source measurements. The device-under-test (DUT) was mounted inside a sample holder thermally anchored to the mixing chamber of the dilution refrigerator.}
    \label{fig:measCircuit}
\end{figure}

\section{The effect of noise on the current-voltage characteristics}

The noise reduction was our primary concern in characterization of low-$E_J$ junctions. Most of our measurements have been performed in the constant current mode. According to Eq. 1, $I_C^{AB}=30$ nA at $T=0$ for an $Al/AlOx/Al$ JJ with $E_J=1$ K. With further reduction of $E_J$ and increase of the phase slip rate, the current range well below 1 nA becomes relevant. 

Figure A3 illustrates the importance of proper filtering of noises in both the current supply part and the voltage recording part of the measuring setup. By using the combination of cascaded low-pass filters and $100 k\Omega$ resistors on the sample holder, we were able to record switching currents in the pA range (Fig. 2 c of the main text.).

\begin{figure}
    \centering
    \includegraphics{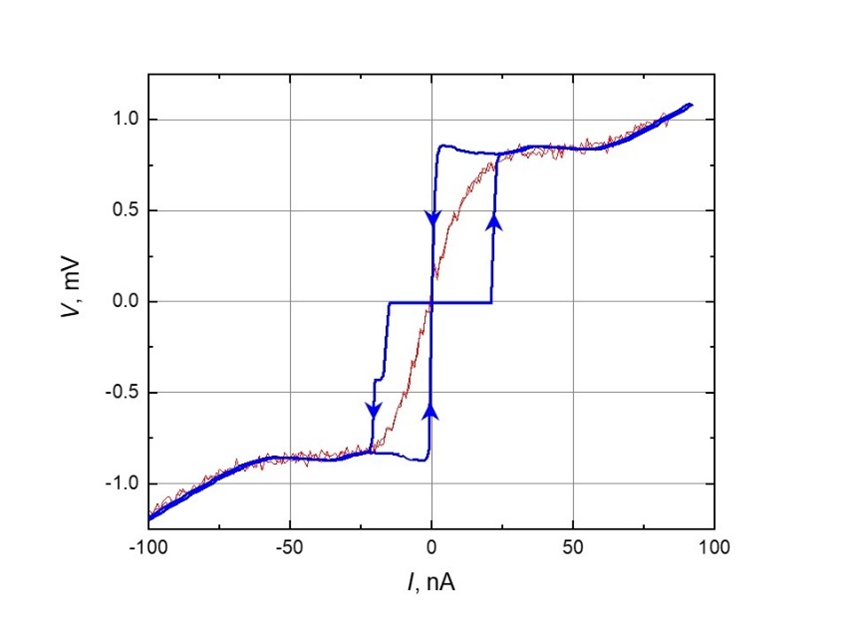}
    \caption{The IVC recorded for a two-unit SQUID device with different measurement setups at $T=25$ mK (sample 3 in Table 1). Without thorough filtering, the IVC was non-hysteretic and smeared. Proper filtering of all leads enables observation of a well-developed hysteresis expected for an underdamped junction at low $T$.}
    \label{fig:my_label}
\end{figure}

\section{Modeling the effect of thermal noise}

The theory of the DC transport in underdamped Josephson junctions in presence of a stochastic noises has been developed by Ivanchenko and Zilberman  \cite{Ivanchenko1969-et}. According to the Ivanchenko-Zilberman (IZ) model, a voltage-biased Josephson junction is subject to thermal noise of the biasing resistor which causes a phase diffusion (see Eq. 4 in the main text). By solving the corresponding Fokker-Planck equation, the superconducting part of the current as a function of bias voltage $V_B$ can be found as:

\begin{equation}
    I_S = I_C \times Im[\frac{J_{1-i\alpha \nu}(\alpha)}{J_{i\alpha \nu}(\alpha)}],
\end{equation}

where $\alpha=\frac{E_J}{k_BT}$, $\nu=\frac{V_B}{I_CR}$ and $J_{a+ib}$ is the modified Bessel function. In the limit of a small Josephson energy $E_J\ll k_B $ this expression is simplified: 

\begin{equation}
    I_S = \frac{I_CR}{2} \frac{V_B}{V_B^2+V_n^2} 
\end{equation}
\begin{equation}
    V_n = \frac{2e}{\hbar}Rk_BT
\end{equation}

The maximum current that can be carried by Cooper pairs is realized at $V_B = V_n$ ($V_n$ is the Johnson noise from the resistor $R$); the further increase of the biasing current leads to switch to the resistive state. As Fig. A4.b shows, the theoretical value of the switching current predicted by the classical IZ (cIZ) model starts to deviate from $I_C^{AB}$ when the thermal fluctuations exceed the Josephson energy. The maximum value of the switching current is realized at a non-zero voltage $V_n$, which depends only on the voltage noise amplitude, so the zero-bias resistance in the IPS regime is expected to scale as $E_J^{-2}$ .

\begin{figure}
    \centering
    \includegraphics[scale=0.65]{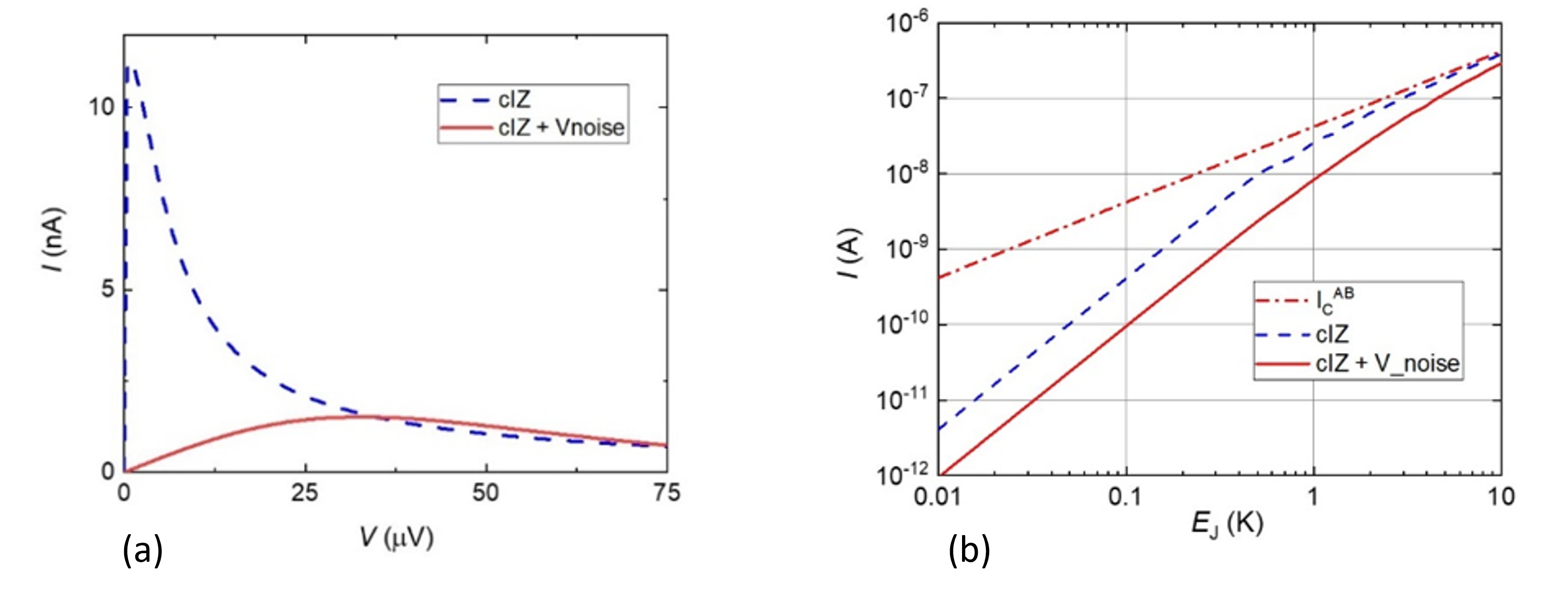}
    \caption{(a) The supercurrent branch of the IVC of a JJ with $E_J= 1$ K at $T=50$ mK, predicted by the classical IZ theory (dashed line) and its modification in presence of gaussian noise with amplitude $24\mu V$ (solid line).  (b) Nominal critical current (dot-dashed line), switching current according to the IZ theory without (dashed line) and with (solid line) extra voltage noise of the same amplitude.}
    \label{fig:modelNoise}
\end{figure}

In the case when a system is subject to other sources of noise such as the thermal noise across the junction capacitance or the external electromagnetic noise due to insufficient filtering, the modified IVC can be calculated by convolving the cIZ curve with Gaussian-distributed $V_B$ of the width corresponding to the noise amplitude $V_{noise}$ (Figs. A4.a and A4.b). As Fig. A4.b shows, the cIZ model can explain qualitatively the switching current behavior in systems with low Josephson energy, the value of the excessive noise $V_{noise}$ could be used as a fitting parameter to obtain quantitative agreement. 

\thanks{Footnote to title of article.}

\section{Supplemental References}
%

\email{gersh@physics.rutgers.edu}
\affiliation{Physics Department, Rutgers University}